\pdfoutput=1

\documentclass[%
 reprint,
 showpacs,
 nofootinbib,
 amsmath,amssymb,
 aps,
 pra,
 longbibliography
]{revtex4-1}

\usepackage{graphics}
\usepackage{graphicx}
\usepackage{color}
\usepackage{amsmath}
\usepackage{slashed}
\usepackage{epsfig}
\usepackage{hyperref}

\usepackage{color}

\newcommand{\eq}[1]{Eq.~(\ref{#1})}
\newcommand{\fig}[1]{Fig.~\ref{#1}}
\newcommand{\tab}[1]{Table~\ref{#1}}
\renewcommand{\sec}[1]{Sec.~\ref{#1}}
\newcommand{\citeref}[1]{Ref.~\cite{#1}}
\newcommand{\citerefs}[1]{Refs.~\cite{#1}}

\usepackage{titlesec}
\titlespacing{\section}{0pt}{1.5ex plus .2ex minus .2ex}{3ex plus .2ex}
\titleformat{\section}[runin]
       {\normalfont\bfseries\itshape}
       {\thesection.}
       {0.5em}
       {}
       [:]

\begin{document}

\title{$\nu$DFSZ: a technically natural non-supersymmetric model \\ of neutrino masses,
baryogenesis, the strong CP problem, and dark matter}

\author{Jackson D. Clarke and Raymond R. Volkas}
\affiliation{ARC Centre of Excellence for Particle Physics at the Terascale, \\ 
School of Physics, University of Melbourne, VIC 3010, Australia.}

\date{\today}

\begin{abstract}

We describe a minimal extension of the standard model by
three right-handed neutrinos, a scalar doublet, and a scalar singlet
(the ``$\nu$DFSZ'') which serves as an existence proof that
weakly coupled high-scale physics can naturally explain phenomenological shortcomings of the SM.
The $\nu$DFSZ can explain neutrino masses, baryogenesis, the strong $CP$ problem, and dark matter,
and remains calculably natural despite a hierarchy of scales up to $\sim 10^{11}$~GeV.
It predicts a SM-like Higgs boson, (maximally) TeV-scale scalar states, 
intermediate-scale hierarchical leptogenesis ($10^5\text{ GeV}\lesssim M_{N}\lesssim 10^7\text{ GeV}$),
and axionic dark matter.

\end{abstract}


\maketitle

\section{Introduction}

The standard model (SM) and the paradigm of
spontaneous electroweak symmetry breaking, realised by a scalar potential
\begin{align}
 V_{SM} = \mu^2\Phi^\dagger\Phi + \lambda (\Phi^\dagger\Phi)^2,
\end{align}
where $\mu^2(m_Z) \approx -(88\text{ GeV})^2$,
has proven extremely successful in explaining
low energy phenomena.
Nevertheless, it fails to explain neutrino masses,
the baryon asymmetry of the Universe (BAU),
the smallness of the neutron electric dipole moment (the strong $CP$ problem), 
dark matter, and gravity.
Whether nature realises these phenomena
in a ``natural'' way, i.e.,
in such a way that $\mu^2$ is
(sufficiently) insensitive to physically meaningful quantum corrections,
remains an open question.
Still, motivated by aesthetics,
the pursuit of a natural ``theory of everything'' has motivated
much of modern particle physics.

In the same vein, this paper describes an extension of the SM by
three right-handed neutrinos, a scalar doublet, and a scalar singlet.
The model can be thought of as an extension of the 
Dine--Fischler--Srednicki--Zhitnitsky (DFSZ) invisible axion model \cite{Dine1981rt,Zhitnitsky1980tq}
by right-handed neutrinos, and is thus
henceforth referred to as the $\nu$DFSZ.
Notably, as we will describe in this paper, the $\nu$DFSZ can explain
neutrino masses, the BAU, the strong $CP$ problem, and dark matter
\textit{in a calculably natural way},
even despite a hierarchy of scales up to $\sim 10^{11}$~GeV.
This is achieved by a seesaw mechanism, hierarchical leptogenesis, 
the Peccei--Quinn (PQ) mechanism, an invisible axion, 
and a technically natural decoupling limit, respectively.

The paper is organised as follows.
We first detail the $\nu$DFSZ, its vacuum, and its scalar sector (and constraints).
We then describe how it provides explanations for the strong $CP$ problem, 
dark matter, neutrino masses, and the BAU.
Penultimately, we describe our naturalness philosophy,
identify the symmetries which protect each scale from quantum corrections,
and study an example point in the parameter space.
Finally we conclude.

\section{The $\nu$DFSZ Lagrangian\label{secnudfsz}}

The scalar content of the model is a complex singlet $S$ and
two complex doublets $\Phi_{1,2}$ of hypercharge +1.
The potential is\footnote{As 
far as we are aware, the $\nu$DFSZ was first discussed in \citeref{Volkas1988cm}.
See Refs.~\cite{Bertolini2014aia,Dias2014osa,Dias2006th} and references therein for other
minimal approaches to connecting the PQ mechanism with neutrino masses.
}
\begin{align}
 V_{\nu\text{DFSZ}} =&\,
  M^2_{11}\, \Phi_1^\dagger \Phi_1
  + M^2_{22}\, \Phi_2^\dagger \Phi_2 
  +M_{SS}^2\, S^*S
  \nonumber\\ &
  + \frac{\lambda_1}{2} \left( \Phi_1^\dagger \Phi_1 \right)^2
  + \frac{\lambda_2}{2} \left( \Phi_2^\dagger \Phi_2 \right)^2
  + \frac{\lambda_S}{2}\left(S^*S\right)^2
  \nonumber\\ &
  + \lambda_3\, \left(\Phi_1^\dagger \Phi_1\right) \left(\Phi_2^\dagger \Phi_2\right)
  + \lambda_4\, \left(\Phi_1^\dagger \Phi_2\right) \left(\Phi_2^\dagger \Phi_1\right)
  \nonumber\\
    &   +\lambda_{1S}\left(\Phi_1^\dagger\Phi_1\right)\left(S^*S\right)
        +\lambda_{2S}\left(\Phi_2^\dagger\Phi_2\right)\left(S^*S\right) \nonumber \\
    &   -\epsilon\Phi_1^\dagger\Phi_2S^2
        -\epsilon\Phi_2^\dagger\Phi_1S^{*2}, \label{EqnuDFSZ}
\end{align}
where $M_{SS}^2\sim -\left(10^{11}\text{ GeV}\right)^2 \equiv -M_{PQ}^2$ 
sets the PQ scale.
Additional terms otherwise allowed by gauge symmetry 
are forbidden by a global $U(1)_{PQ}$ symmetry
to be defined in \sec{secstrongcp},
which is essential in solving the strong $CP$ problem.
The $\epsilon$ terms are necessary\footnote{Another
option is to have terms $-(\kappa \Phi_1^\dagger\Phi_2 S +H.c.)$ \cite{Langacker1986rj}.}
to assign a PQ charge to $S$
and help to generate neutrino masses
once $S$ obtains a vacuum expectation value (vev).

The only addition to the SM fermionic content is three right-handed neutrinos.
The strong $CP$ solution
dictates that $\Phi_1$ ($\Phi_2$) couple to $u_R$ ($d_R$),
and our solution for natural neutrino masses and leptogenesis
requires that $\Phi_2$ couple to $\nu_R$.
The Yukawa Lagrangian is therefore
\begin{align}
  -\mathcal{L}_Y =&\, + y_u\overline{q_L}\tilde\Phi_1 u_R + y_d\overline{q_L}\Phi_2 d_R \nonumber \\
   & + y_e \overline{l_L}\Phi_J e_R + y_\nu \overline{l_L}\tilde\Phi_2 \nu_R \nonumber \\
   & + \frac12 y_N \overline{(\nu_R)^c}S\nu_R + H.c. , \label{EqYuk}
\end{align}
where family indices are implied, $\tilde{\Phi}_i\equiv i\tau_2\Phi_i^*$,
and $J=2\ (1)$ is a Type II (Flipped) $\nu$-two-Higgs-doublet model ($\nu$2HDM) arrangement.
We will work in the basis where $y_N$ is diagonal and real.
Again, additional terms are forbidden by the $U(1)_{PQ}$ symmetry.\footnote{In 
this model the right-handed neutrinos gain mass from the vev of $S$,
but an alternative scenario with explicit Majorana masses is also possible.}

We note here that each of $\epsilon \to 0$, $y_N \to 0$, and $y_\nu \to 0$
is a technically natural limit, since they lead to an extra
$U(1)$ symmetry which can be identified with lepton number.
As well, there are two apparent technically natural decoupling limits
associated with enhanced Poincar\'e symmetries \cite{Foot2013hna}:
$\epsilon, \lambda_{1S}, \lambda_{2S}, y_N \to 0$ decouples $S$, and
$\epsilon, \lambda_{1S}, \lambda_{2S}, y_\nu \to 0$ decouples the $(\nu_R,S)$ subsystem.
These limits will prove important in protecting the
hierarchy of scales in the model.

\section{Vacuum}

The scalar fields acquire vevs
$\langle S \rangle \equiv v_S/\sqrt{2}$,
$\langle \Phi_i \rangle \equiv \left(0,v_i/\sqrt{2}\right)^T$.
If $v_S \gg v_i$, then
\begin{align}
 v_S \approx \sqrt{\frac{-2M_{SS}^2}{\lambda_S}}\sim 10^{11}\text{ GeV}.
\end{align}
Expanding around this vev,
the right-handed neutrinos acquire Majorana masses $M_N=y_N\langle S \rangle$
and the scalar potential becomes
\begin{align}
 V_{\nu\text{2HDM}} \approx&\;
m^2_{11}\, \Phi_1^\dagger \Phi_1
+ m^2_{22}\, \Phi_2^\dagger \Phi_2 \nonumber\\ &
-m^2_{12}\, \left(\Phi_1^\dagger \Phi_2 + \Phi_2^\dagger \Phi_1\right) + ... \ , \label{EqV2HDM}
\end{align}
where $m^2_{ii} = M^2_{ii}+\lambda_{iS}\langle S \rangle^2$ and
$m^2_{12} = \epsilon\langle S \rangle^2$.

We will adopt the natural explanation of neutrino masses and baryogenesis
detailed in \citeref{Clarke2015hta}.
This requires $v_2 \sim \mathcal{O}(1$--10)~GeV
achieved with $m_{11}^2<0$, $m_{22}^2>0$, and $m_{12}^2/m_{22}^2\ll 1$.\footnote{Note 
that, like $\epsilon\to 0$ in the original Lagrangian,
$m_{12}^2/m_{22}^2\to 0$ is a technically natural limit 
associated with $U(1)_L$ \cite{Ma2000cc}.}
Anticipating $m_{22}^2\gg v_1^2(\lambda_{3}+\lambda_4)$, $\lambda_2 v_2^2$, 
the $\Phi_{i}$ vevs can be written
\begin{align}
 v_2 \equiv \frac{v_1}{\tan\beta} \approx \frac{m_{12}^2}{m_{22}^2}v_1 , &&
 v_1 \approx \sqrt{\frac{2}{\lambda_1}\left( -m_{11}^2 + \frac{ m_{22}^2 }{\tan^2\beta} \right)},
\end{align}
where $\sqrt{v_1^2+v_2^2}=v\approx 246$~GeV
and we have implicitly defined $\tan\beta$.
There is an important consistency condition
$2 m_{22}^2/\tan^2\beta \lesssim \lambda_1 v_1^2$
to ensure $m_{11}^2<0$ 
and avoid a fine-tuning for $v$ (see \citeref{Clarke2015hta}).

Typical values for the mass parameters are $m_{11}^2 \approx -(88\text{ GeV})^2$,
$m_{22}\sim 10^3$~GeV, and $m_{22}^2/\tan^2\beta \ll |m_{11}^2|$.
Therefore, for no fine-tuning between $M_{ii}^2$ and $m_{ii}^2$,
we already expect $\lambda_{1S}\lesssim 10^{-18}$,
$\lambda_{2S}\lesssim 10^{-16}$, and $\epsilon \ll 10^{-18}$.

\section{Scalar sector}

The scalar mass eigenstates are, 
up to $v_1/m_{22}$ and $m_{12}^2/m_{22}^2$ corrections 
(see \citeref{Clarke2015hta} for expressions),
a $CP$ even state ($h$) with $m_{h}^2\approx \lambda_1 v_1^2$,
three heavy scalar states ($H,A,H^\pm$) with masses $\approx m_{22}$,
a PQ-scale neutral scalar ($s$) with $m_s^2=\lambda_S v_S^2$,
and a very light pseudo-Goldstone boson (the invisible axion).

\begin{figure}[t]
 \centering
 \includegraphics[width=0.9\columnwidth]{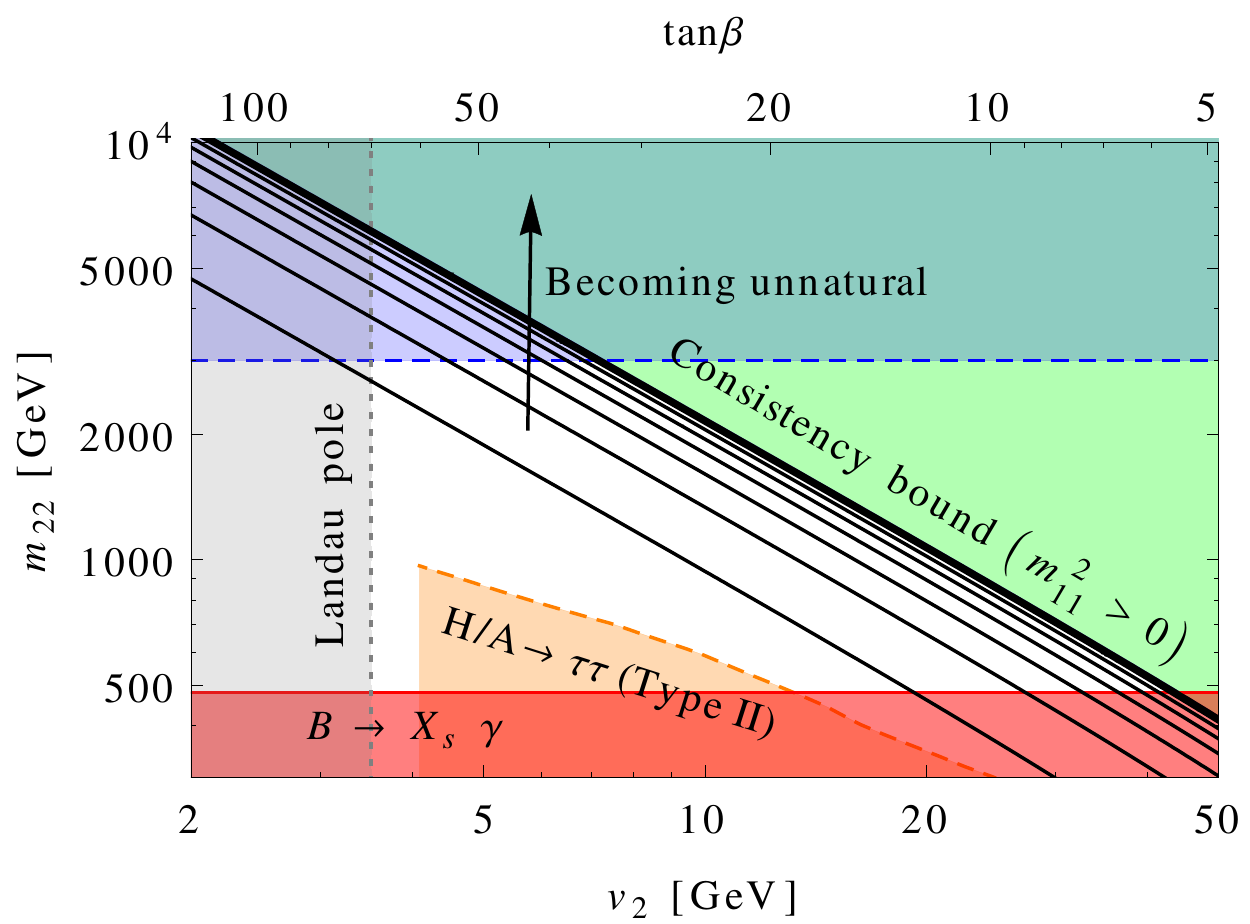}
 \caption{Constraints on $m_{22}$, as labelled.
 Solid black contours are $m_{11}^2/\text{GeV}^2=-80^2,-70^2,$ and so on.
 The $H/A\to \tau\tau$ bound is taken from the CMS search \cite{Khachatryan2014wca}.
 The naturalness bound is only illustrative (see \citeref{Clarke2015hta}).}
 \label{FigScalarCons}
\end{figure}

Owing to the approximate $U(1)$ symmetry due to
$m_{12}^2/m_{22}^2\ll 1$ and $v^2/m_{22}^2\ll 1$,
the state $h$ closely resembles the SM Higgs.\footnote{Up 
to $v/v_S$ corrections, at low scale the model is
just the $\nu$2HDM discussed in \citeref{Clarke2015hta}
with a very weakly coupled axion.}
In \fig{FigScalarCons} we illustrate the various constraints on $m_{22}$.
These are the aforementioned consistency condition,
measurements of $B\to X_s\gamma$ \cite{Amhis2014hma,Misiak2015xwa},
$H/A\to \tau\tau$ LHC searches (for the Type II model) \cite{Aad2014vgg,Khachatryan2014wca},
perturbativity up to the Planck scale \cite{Bijnens2011gd},
and naturalness \cite{Clarke2015hta}.
The naturalness bound was derived in Ref.~\cite{Clarke2015hta}
subject to the naturalness condition we describe in Sec.~\ref{secnatphilo},
and we refer the reader there for details.

\section{The strong $CP$ problem\label{secstrongcp}}

Gauge invariance and renormalisability permit the
physical $CP$ violating term,
$\bar{\theta} \frac{\alpha_s}{8\pi}G_a^{\mu\nu}\tilde G^a_{\mu\nu}$
where $G$ is the gluon field strength tensor and $\tilde G$ its dual,
to be added to the SM Lagrangian.
Bounds on the neutron electric dipole moment constrain $\bar{\theta}\lesssim 10^{-10}$ \cite{Baker2006ts}.
The strong $CP$ problem is: why is $\bar{\theta}\approx 0$?\footnote{For
reviews of the strong $CP$ problem and axions see e.g. Refs.~\cite{Peccei2006as,Raffelt2006cw,Sikivie2006ni,Kim2008hd,Agashe2014kda}.}

The PQ solution \cite{Peccei1977hh} to the strong $CP$ problem
is to demand a global chiral $U(1)_{PQ}$ symmetry;
if the sum of the $u_R$ and $d_R$ PQ charges is non-zero then, 
after PQ symmetry breaking, $\bar{\theta}$ becomes dynamical, and
the vacuum potential ``selects'' $\langle\bar{\theta}\rangle = 0$.
The $\nu$DFSZ model is one manifestation of this solution.

Let us now identify the global $U(1)_{PQ}$ symmetry.
Defining the $U(1)_{PQ}$ charge names as in \tab{TabPQ},
we can (without loss of generality) set $X_q = 0$ and $X_u+X_d = 1$.
Equations (\ref{EqnuDFSZ}) and (\ref{EqYuk}) 
set an additional six constraints on the charges,
which brings the total to eight for nine unknown charges.
They are completely defined by setting
$X_1 = \alpha X_2$, as long as $\alpha\ne 1$;
it is convenient to choose $\alpha = -\cot^2\beta$ 
so that the PQ current does not couple to the field
eaten by the $Z$ boson.
The resulting charge values are given in \tab{TabPQ}.

\begin{table}[t]
 \begin{tabular}{c|c|c}
   \hline\hline
   Field	& $U(1)_{PQ}$ & Value in $\left[\text{Type II, Flipped}\right]$ \\
   \hline
   $S$		& $X_S$ 	& $\frac12$		 \\
   $\Phi_1$	& $X_1$ 	& $\cos^2\beta$	 \\
   $\Phi_2$	& $X_2$ 	& $-\sin^2\beta$	 \\
   $q_L$	& $X_q$ 	& $0$			 \\
   $u_R$	& $X_u$ 	& $\cos^2\beta$	 \\
   $d_R$	& $X_d$ 	& $\sin^2\beta$	 \\
   $l_L$	& $X_l$ 	& $\frac34-\cos^2\beta$ \\
   $\nu_R$	& $X_\nu$ 	& $-\frac14$		 \\
   $e_R$	& $X_e$ 	& $\left[\frac74,\frac34\right]-2\cos^2\beta$ \\
   \hline\hline
 \end{tabular}
 \caption{Charges of fields under the PQ symmetry.}
 \label{TabPQ}
\end{table}

A final comment before moving on.
In the SM, if $\bar{\theta}$ is set to zero by hand at some high scale,
renormalisation implies $\bar{\theta}\lesssim 10^{-17}$ \cite{Ellis1978hq,Georgi1986kr},
well below the experimental bound.
In this sense, in the SM, $\bar{\theta}\approx 0$ is \textit{already} natural.
Yet this explanation remains unsatisfying,
since the limit $\bar{\theta}\to 0$
cannot be identified with a symmetry.
The $\nu$DFSZ solution requires $\lambda_{iS}\lll 1$,
and thus one could similarly ask: why are the $\lambda_{iS}\approx 0$?
At least, here, this limit is identified with an 
increased Poincar\'e symmetry.
As well, in the presence of $CP$ violating new physics 
(such as the right-handed neutrinos),
this solution \textit{guarantees} $\bar{\theta}\approx 0$.

\section{Dark matter\label{secdarkmatter}}

The $\nu$DFSZ axion gains a mass
\begin{align}
 m_a \approx 6\text{ $\mu$eV} \left( \frac{10^{12}\text{ GeV}}{f_a} \right) \label{Eqaxionmass}
\end{align}
due to the chiral anomaly \cite{Raffelt2006cw,Kim2008hd},
where $f_a\approx \langle S \rangle$ is the axion decay constant,
and inherits $v/f_a$ suppressed 
couplings to nucleons, photons, and electrons 
(for expressions see e.g. \citeref{Raffelt2006cw}).
Stellar energy loss 
constrains $f_a\gtrsim 4\times 10^8$~GeV \cite{Raffelt2006cw}.\footnote{In 
a Type II $\nu$DFSZ,
red giants and white dwarfs constrain $f_a\gtrsim 8\times 10^{8}\sin^2\beta$~GeV 
(the white dwarf cooling fit is actually improved for $f_a\approx 1\times 10^9 \sin^2\beta$) 
\cite{Raffelt1985nj,Raffelt2006cw,Raffelt1994ry,Isern2008nt}.}

The axion provides a cold dark matter candidate via the
misalignment mechanism \cite{Preskill1982cy,Abbott1982af,Dine1982ah},
wherein a significant amount of energy density is stored
in coherent oscillations of the axion field, \cite{Sikivie2006ni}
\begin{align}
 \Omega_a h^2 \sim 0.7\left(\frac{f_a}{10^{12}\text{ GeV}}\right)^\frac76
  \left(\frac{\theta}{\pi}\right)^2,
\end{align}
where $-\pi\le\theta\le\pi$ is the misalignment angle.
The requirement that this quantity not exceed the measured cold dark matter energy density 
$\Omega_{\text{CDM}}h^2 \approx 0.12$ \cite{Planck2015xua} implies
\begin{align}
 f_a \lesssim 2\times 10^{11}\text{ GeV} \left( \frac{\pi}{\theta} \right)^\frac{12}{7},
\end{align}
with equality reproducing the observed density.
If the PQ symmetry is broken after inflation, 
then the misalignment angle is the average value taken over many distinct patches,
$\theta^2\approx \pi^2/3$, and one obtains $f_a\lesssim 6\times 10^{11}$~GeV \cite{Peccei2006as}.\footnote{If 
inflation occurred after PQ symmetry breaking
then a smaller $\theta$ can be
anthropically chosen, allowing $f_a>10^{12}$~GeV \cite{Pi1984pv}.}
Future projections of the ADMX and CAPP resonant microwave cavity experiments 
promise to probe this interesting region of parameter space \cite{Sikivie1983ip,CAPP}.

\section{Neutrino masses\label{secnumasses}}

The neutrino mass matrix is given by
\begin{align}
 m_\nu = \frac{v_2^2}{2} y_\nu \mathcal{D}_M^{-1} y_\nu^T 
   \approx \frac{1}{\tan^2\beta} \left( \frac{v^2}{2} y_\nu \mathcal{D}_M^{-1} y_\nu^T \right), \label{Eqnumass2HDM}
\end{align}
where the bracketed quantity is the 
typical Type I seesaw formula \cite{Minkowski1977sc,Mohapatra1979ia,Yanagida1979as,GellMann1980vs}.
The mass matrix is diagonalised by a unitary matrix $U$,
$U m_\nu U^T = \text{diag}(m_1,m_2,m_3) \equiv \mathcal{D}_m$,
where $m_i$ are the neutrino masses.
It will be useful to express $y_\nu$
in the Casas--Ibarra \cite{Casas2001sr} form,
\begin{align}
 U y_\nu = \frac{\sqrt{2}}{v_2} \mathcal{D}_m^\frac12 R \mathcal{D}_M^\frac12, \label{EqCassasIbarra}
\end{align}
where $R$ is a (possibly complex) orthogonal matrix.

\section{The BAU\label{secbau}}

The BAU is produced analogously to standard
hierarchical thermal leptogenesis \cite{Fukugita1986hr}, 
via the out-of-equilibrium, $CP$ violating decays 
of the lightest right-handed neutrino: $N_1 \to l \Phi_2$ \cite{Clarke2015hta}.

\begin{figure}[t]
 \centering
 \includegraphics[width=0.9\columnwidth]{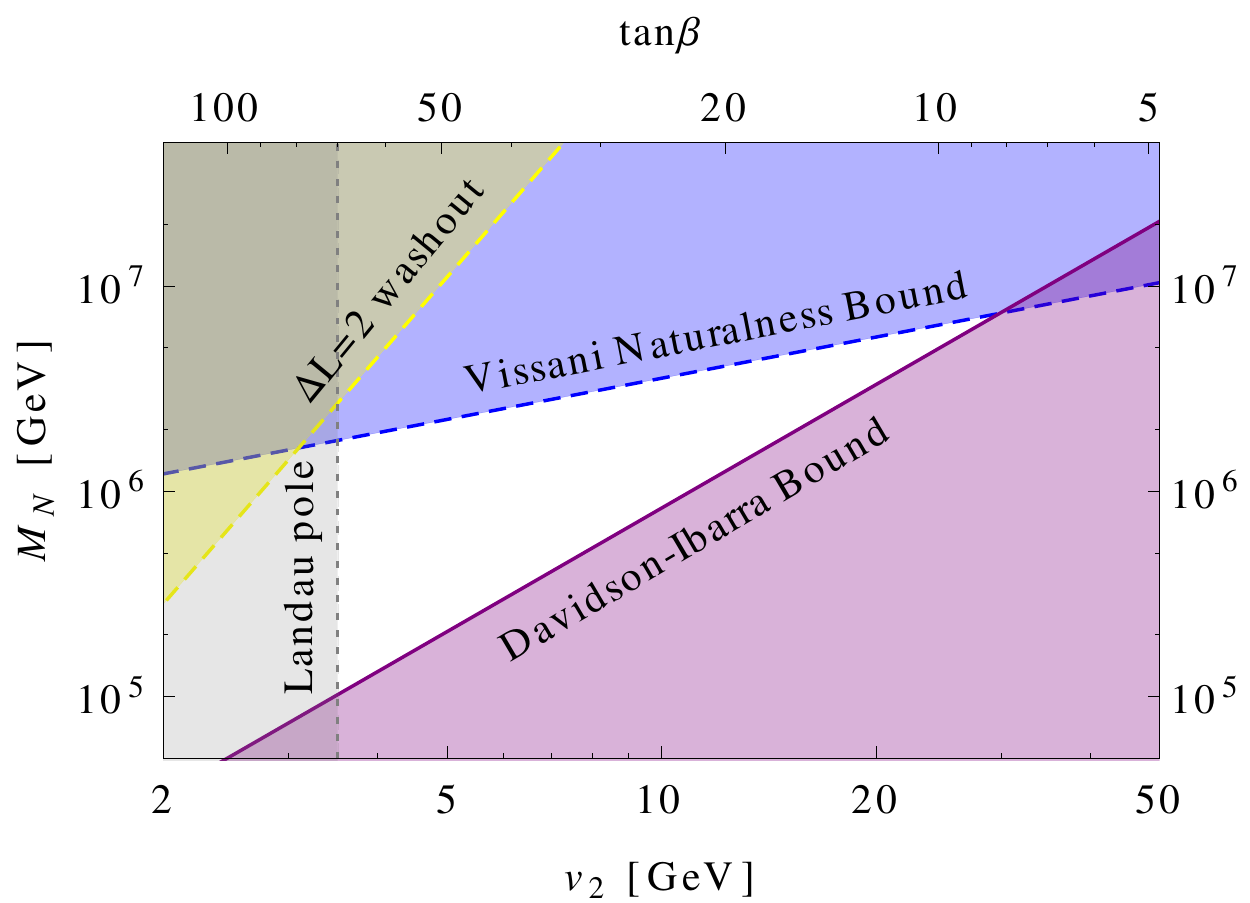}
 \caption{Constraints on the right-handed neutrino masses.
 The naturalness bound on $M_N$ corresponds to the rough bound \eq{EqMNNatural}
 evaluated at $m_{22}=1$~TeV.}
 \label{FigNaturalMN}
\end{figure}

If only decays and inverse decays are considered,
the evolution of the asymmetry is characterised 
(in the one-flavour approximation)
by the decay parameter $K$,
\begin{align}
 K = \frac{\Gamma_D}{H|_{T=M_1}} \equiv \frac{\tilde m_1}{m_*},
\end{align}
where $\Gamma_D$ is the $N_1$ decay rate, 
$H$ is the Hubble rate,
$\tilde m_1$ is the effective neutrino mass,
and $m_*$ is the equilibrium neutrino mass,
\begin{align}
 \Gamma_D & = \frac{1}{8\pi} (y_\nu^\dagger y_\nu)_{11} M_1, &
 \tilde m_1 & \equiv \frac{ (y_\nu^\dagger y_\nu)_{11} v_2^2}{2 M_1}, \\
 H & \approx \frac{ 17 T^2 }{M_{Pl}}, &
 m_* & \approx \frac{ 10^{-3} \text{ eV} }{\tan^2\beta} . \label{Eqmtildemstar}
\end{align}
The salient $\Delta L=1$ scatterings are electroweak, and those
involving $b$ quarks and (in Type II) $\tau$ leptons.
Since those rates scale with the decays and inverse decays,
proportional to $(y_\nu^\dagger y_\nu)_{11}$,
they can only represent a minor correction to the standard
case with electroweak and $t$ quark scatterings.
The $\Delta L =2$ scatterings \textit{can} however constitute
a significant departure from the standard case,
particularly in $K\ll 1$ scenarios
dependent on initial conditions.
For the parameter space of interest to us,
the generated asymmetry is safe 
from strong $\Delta L =2$ scattering washout,
as shown in \fig{FigNaturalMN} \cite{Clarke2015hta}.
As well, since leptogenesis in this model will be
occurring at temperatures below $10^9$~GeV,
the Yukawa couplings will be in equilibrium
and flavour effects cannot be ignored 
(see e.g. Refs.~\cite{Abada2006fw,Nardi2006fx,Abada2006ea,Blanchet2006be,Dev2014laa}).\footnote{An 
additional consideration is $N_1N_1\to aa$ annihilations.
\citeref{Langacker1986rj} estimates
$\Gamma_{N_1N_1\to aa} \sim 10^{-2} M_{N_1}^5/\langle S \rangle$
at $T=M_{N_1}$, which implies the out-of-equilibrium condition $M_{N_1}\lesssim 10^9$~GeV,
easily satisfied the parameter space of interest.}

These departures from the standard scenario deserve further detailed study.
Still, we do not expect the picture to be dramatically changed.
In particular we expect the 
Davidson--Ibarra bound \cite{Davidson2002qv,Giudice2003jh}
for successful hierarchical thermal leptogenesis,
scaled for the differing vev in \eq{Eqnumass2HDM},
to approximately hold:
\begin{align}
 M_{N_1} \gtrsim \frac{5\times 10^8\text{ GeV}}{\tan^2\beta}.
\end{align}
This bound is depicted in \fig{FigNaturalMN}.

Let us briefly demonstrate that this picture is consistent.
A simple example configuration which achieves 
maximal $CP$ violation and saturates the Davidson--Ibarra bound is
(assuming normal ordering)
$m_1\ll \tilde{m}_1$ and
\begin{align}
  R = \left(
 \begin{array}{ccc}
  \sqrt{1-R_{31}^2} 	& 0 		& R_{31} \\
  0			& 1	 	& 0 \\
  R_{31} 		& 0 		& -\sqrt{1-R_{31}^2}
 \end{array}
 \right) , \label{EqRMatrix}
\end{align}
where $R_{31}\equiv \kappa \exp(i\pi/4)$.
Here $\kappa$ is related to the decay parameter by
\begin{align}
 \kappa \approx \frac{0.15\sqrt{K}}{\tan\beta}\left(\frac{0.05\text{ eV}}{m_3} \right)^\frac12, \label{EqKappa}
\end{align}
and is typically $\lesssim 10^{-2}$ in the parameter range of interest.
In the limit $m_1=0$, 
this corresponds to a $U y_\nu$ which has one zero row,
but is otherwise approximately diagonal.
We note that, in this configuration, 
$\tilde{m}_1$ and the $CP$ asymmetry
are sufficiently stable under radiative corrections.
The reader may check this claim against the
renormalisation group equation (RGE) in Appendix~\ref{AppRGEs}.

\section{Our naturalness philosophy\label{secnatphilo}}

A naturalness problem arises when a low mass scale
is subject to large and physically meaningful quantum corrections
from a high mass scale.
In particular, the electroweak scale can
receive such corrections from high-scale new physics.

The RGE formalism is a sensible way to quantify a naturalness problem
in any perturbative quantum field theory.
For example, in the SM, the $\mu^2(\mu_R)$ RGE is
dominated by the top quark Yukawa,
\begin{align}
 \frac{d\mu^2}{d\ln\mu_R} \approx \frac{1}{(4\pi)^2} 6 y_t^2 \mu^2 ,
\end{align}
where $\mu_R$ is a representative energy scale.
From the RGE perspective, 
$\mu^2$ is not subject to any large physical corrections
and can therefore be considered natural,
in the sense that once it is set to be electroweak scale 
it stays as such within the region of validity of the model ($\mu_R < M_{Pl}$).\footnote{It still may 
be the case that quantum gravity effects at the Planck scale induce a naturalness problem.  
However, this cannot be rigorously computed in the absence of a cogent theory of
quantum gravity, so an agnostic stance on this possible problem seems reasonable to us.
In other words, we remain agnostic to the physics which
sets the boundary condition $\mu^2(\mu_R\ggg m_Z)\lesssim$~TeV$^2$ at some high scale;
this is a separate problem -- a hierarchy-type problem -- which
may or may not have a natural solution
(see e.g. \citerefs{Bardeen1995kv,Lykken2013,Tavares2013dga} for relevant discussion).
The naturalness question induced by the high PQ scale can, by contrast, be fully analysed
within quantum field theory, and we limit our scope to that.}
It follows intuitively that $\mu^2(\mu_R\ggg m_Z)$ is not finely tuned against $\mu^2(m_Z)$.
In the following we demonstrate that
there exists a region of $\nu$DFSZ parameter space 
where our phenomenological goals can be achieved
and the heavy PQ scale induces no naturalness problem,
i.e., all scales remain stable under RG evolution.

\section{Naturalness in the $\nu$DFSZ\label{secnatnudfsz}}

Defining $\mathcal{D}\equiv (4\pi)^2\frac{d}{d\ln\mu_R}$ and keeping only $y_{t,b,\tau,\nu}$ Yukawas,
the one-loop RGEs for the [Type II, Flipped] $\nu$2HDM mass parameters can be written \cite{Lyonnet2013dna}
\begin{widetext}
\begin{align}
 \mathcal{D} m_{12}^2 = &\
   m_{12}^2 \left( -\frac32 g_1^2 -\frac92 g_2^2 
 + 2\lambda_3 + 4\lambda_4 + 2\lambda_S + 4\lambda_{1S} + 4\lambda_{2S}
 + 3y_t^2 + 3y_b^2 + y_\tau^2 + \text{Tr}\left(y_\nu^\dagger y_\nu \right) \right), \\
 \mathcal{D} m_{11}^2 = &\
   m_{11}^2 \left( -\frac32 g_1^2-\frac92 g_2^2 +6\lambda_1 +6y_t^2+\left[0,2y_\tau^2\right]\right)
 + m_{22}^2 \left( 4\lambda_3 + 2\lambda_4  \right)
 + \langle S \rangle^2 \lambda_{1S} \left( 4 \lambda_{1S} + 4 \lambda_S \right) 
 + M_{SS}^2  2\lambda_{1S} , \label{Eqm11sqRGE} \\
 \mathcal{D} m_{22}^2 = &\
   m_{22}^2 \left( -\frac32 g_1^2-\frac92 g_2^2 +6\lambda_2 +6y_b^2 +\left[2y_\tau^2,0 \right]
   +2\text{Tr}\left(y_\nu^\dagger y_\nu\right)\right) 
 + m_{11}^2 \left( 4\lambda_3 + 2\lambda_4  \right) \nonumber\\
 &- 4\text{Tr}\left(y_\nu M_N^2 y_\nu^\dagger \right)
 + \langle S \rangle^2 \lambda_{2S} \left( 4 \lambda_{2S} + 4 \lambda_S \right) 
 + M_{SS}^2  2\lambda_{2S} 
 , \label{Eqm22sqRGE} 
\end{align}
\end{widetext}
where $M_N^2 = y_N^\dagger y_N \langle S\rangle^2$ is the (diagonalised) 
right-handed neutrino mass matrix.
The $\langle S\rangle^2$ and $M_{SS}^2$ terms
correspond to the contribution from the heavy real singlet $s$
in the broken phase.
We provide the full list of RGEs in Appendix~\ref{AppRGEs}.

These RGEs make manifest the decoupling limit
$\epsilon, \lambda_{1S}, \lambda_{2S}, \text{Tr}(y_\nu^\dagger y_\nu y_N^\dagger y_N) \to 0$
which protects the scales from large corrections.
First, corrections to $m_{12}^2$ are proportional to $m_{12}^2$,
reflecting the fact that $\epsilon\to 0$
reinstates a $U(1)_L$ symmetry.
Second, because the parameters $\lambda_{3,4}$ are reintroduced
by gauge loops, $m_{11}^2$ can only be protected from $m_{22}^2$ 
by having $m_{22}^2$ not too much larger than $m_{11}^2$;
it was argued in \citeref{Clarke2015hta} that 
$m_{22}\lesssim \text{few}\times 10^3$~GeV
can accommodate a 10\% fine-tuning measured at $M_{Pl}$.
Third, $m_{22}^2$ is protected from the $M_N$ by (roughly)
$\text{Tr}( y_\nu^\dagger y_\nu y_N^\dagger y_N ) / (4\pi^2) \lesssim m_{22}^2/\langle S\rangle^2$;
this translates to the sufficient condition \cite{Vissani1997ys,Clarke2015gwa}
\begin{align}
 M_{N} \lesssim \frac{ 3\times 10^7\text{ GeV} }{\tan^{\frac23}\beta} \left( \frac{m_{22}}{\text{TeV}} \right)^\frac23, \label{EqMNNatural}
\end{align}
for all the right-handed neutrinos,
illustrated in \fig{FigNaturalMN}.
Last, the $m_{ii}^2$ are protected from the 
PQ scale by (again roughly) $\lambda_{iS}\lesssim m_{ii}^2/\langle S\rangle^2$.
We note that there is a lepton box induced correction to $\lambda_{2S}$;
this correction is also bounded by 
$m_{ii}^2/\langle S\rangle^2$ through \eq{EqMNNatural}.

\section{Explicit example}

As a final demonstration
we thought it illustrative to solve 
the coupled set of RGEs for an explicit example.
We consider $\tan\beta = 30$ and neglect running in the following six quantities:
\begin{align}
M_{SS}^2 =&\ -(10^{11}\text{ GeV})^2, &		M_{N_1} =&\ 6\times 10^5 \text{ GeV},	\nonumber\\
 \langle S \rangle^2 =&\ -M_{SS}^2/\lambda_S, &  M_{N_2} =&\ M_{N_3}=10^6 \text{ GeV}, \nonumber\\
 \lambda_S =&\ 0.26, &				y_N =&\ M_N/\langle S \rangle.
\end{align}
We have taken $M_{N_1}$ at the Davidson--Ibarra bound and
$M_{N_{1,2}}$ below the rough naturalness bound \eq{EqMNNatural}.
We take $y_\nu$ according to Eqs.~(\ref{EqCassasIbarra}), (\ref{EqRMatrix}) and (\ref{EqKappa})
with $K=1$, and neglect running here as well.
A glance at the RGEs in Appendix \ref{AppRGEs} will 
convince the reader that neglecting running 
in these parameters is a good approximation.
Decoupling of the heavy degrees of freedom at $m_s$, $M_{N_i}$, and $m_{22}$ 
should be accompanied by an associated shift
in the $\lambda_j$ parameters,
matching to the effective theory below each threshold. 
However, in practice, since the $\lambda_{iS},y_\nu$ are so small
and $m_{22}$ is not too much larger than $m_Z$,
it makes little numerical difference to implement this shift.
Therefore we evolve the following parameters under
the $\nu$DFSZ RGEs:\footnote{For
definiteness we take a Type II arrangement, but the Flipped arrangement
gives very similar results.}
\begin{align}
 \lambda_3 (m_{22}) =&\ \lambda_4(m_{22}) = 0.02, & \lambda_{1S}(m_s) =&\ 10^{-18},  \nonumber\\
 \lambda_1(m_Z)=&\ \lambda_2(m_Z)=0.26 & \lambda_{2S}(m_s) =&\ 10^{-16}, \nonumber\\
 y_t(m_Z) =&\ 0.96/\sin\beta, & g_1^2(m_Z) =&\ 0.13, \nonumber\\
 y_b(m_Z) =&\ 0.017/\cos\beta, & g_2^2(m_Z) =&\ 0.43, \nonumber\\
 y_\tau(m_Z) =&\ 0.010/\cos\beta, & g_3^2(m_Z) =&\ 1.48. \nonumber
\end{align}
Their evolution is shown in Appendix \ref{AppRGEPlots}.\footnote{We 
note that the parameter $\lambda_1$ tends to run negative,
threatening the stability of the electroweak vacuum;
nevertheless we expect the problem to be no worse than in the SM,
i.e., we expect a metastable vacuum.}

\begin{figure}[t]
 \centering
 \includegraphics[width=0.9\columnwidth]{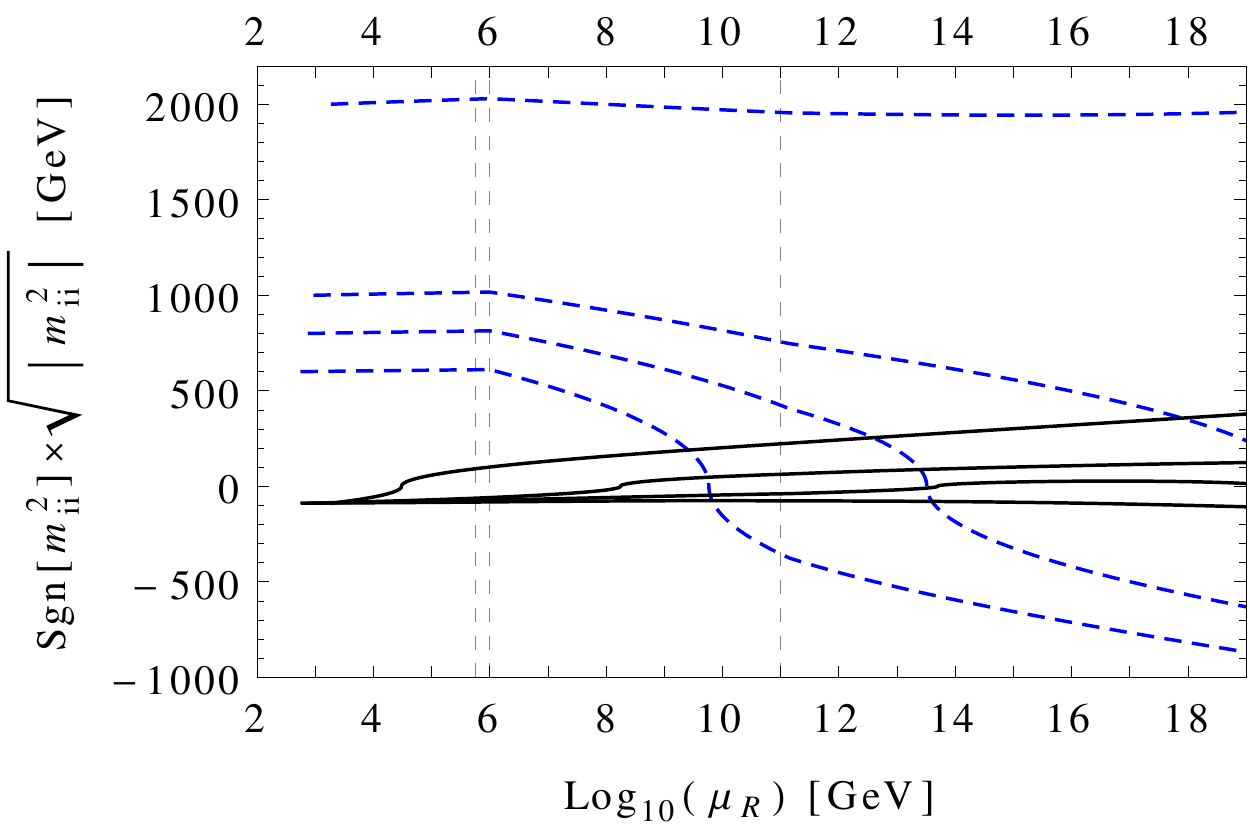}
 \caption{Example RG evolution (see text) of 
 $m_{11}^2(\mu_R)$ (black solid) and $m_{22}^2(\mu_R)$ (blue dashed)
 for $m_{22}=0.6,0.8,1.0,2.0$~TeV bottom to top.}
 \label{FigNaturalnuDFSZ}
\end{figure}

To evolve the mass parameters we set 
$m_{11}^2(m_{22}) = -(88\text{ GeV})^2$
and consider $m_{22}(m_{22}) = 0.6,0.8,1.0,2.0$~TeV.
The $N_i$ and $s$ are decoupled with step functions at their thresholds.
Their RG evolution is shown in \fig{FigNaturalnuDFSZ};
it is plain that the mass parameters in this (viable) example remain relatively small
up to high scales, and are therefore natural according to our philosophy.

\section{Conclusion \label{secconclusion}}

We have described an extension of the SM 
(the ``$\nu$DFSZ'')
by three right-handed neutrinos, 
a complex scalar doublet, and a complex scalar singlet.
The $\nu$DFSZ serves as an existence proof that 
weakly coupled high-scale physics
can explain phenomenological shortcomings of the SM
\textit{without introducing a naturalness problem}.
The model explains 
neutrino masses, the BAU, the strong $CP$ problem, and dark matter,
via a seesaw mechanism, hierarchical leptogenesis, 
the PQ mechanism, and a DFSZ invisible axion, respectively.
It contains four scales:
$|m_{11}|\approx 88$~GeV, $m_{22}\sim 10^3$~GeV,
$M_N\sim 10^5$--$10^7$~GeV, and $M_{PQ}\sim 10^{11}$~GeV,
each protected from quantum corrections by a
technically natural decoupling limit.
The $\sim$TeV-scale scalars and the invisible axion of the model
will be probed in upcoming experiments.

\acknowledgments

This work was supported in part by the Australian Research Council.
We thank Robert Foot for useful input (and anecdotes).

\pagebreak

\bibliography{references}

\begin{widetext}

\appendix

\section{RGEs \label{AppRGEs}}

Following is the full list of RGEs in the [Type II, Flipped] model, found using PyR@TE \cite{Lyonnet2013dna}.
Shown in blue/underlined are those parameters which, for simplicity, we did not evolve in our RGE analysis.
\begin{align}
 \mathcal{D} M_{11}^2 = &\
   M_{11}^2 \left( 6\lambda_1 -\frac32 g_1^2-\frac92 g_2^2 +6y_t^2+\left[0,2y_\tau^2\right]\right)
 + M_{22}^2 \left( 4\lambda_3 + 2\lambda_4  \right)
 + M_{SS}^2  2\lambda_{1S} , \\
 \mathcal{D} M_{22}^2 = &\
   M_{22}^2 \left( 6\lambda_2 -\frac32 g_1^2-\frac92 g_2^2 +6y_b^2 +\left[2y_\tau^2,0 \right]
   +2\text{Tr}\left(y_\nu^\dagger y_\nu\right)\right) 
 + M_{11}^2 \left( 4\lambda_3 + 2\lambda_4  \right)
 + M_{SS}^2  2\lambda_{2S},  \\
\textcolor{blue}{\underline{\mathcal{D} M_{SS}^2}} = &\ 
   M_{SS}^2 \left( 4\lambda_S + \text{Tr}\left(y_N^\dagger y_N\right)\right)
 + M_{11}^2  4\lambda_{1S} 
 + M_{22}^2  4\lambda_{2S} , \\
\textcolor{blue}{\underline{\mathcal{D} \langle S \rangle^2}} = &\ 
  -\text{Tr}\left( y_N^\dagger y_N \right) \langle S \rangle^2 
  \;\;\; \text{[i.e. the wave function renormalisation]}, \\
\mathcal{D} g_{\{1,2,3\}} = &\ \{7,-3,-7\} g_{\{1,2,3\}}^3 , \\
\mathcal{D} \lambda_1 = &\ 
   \frac34 g_1^4 + \frac32 g_1^2g_2^2 + \frac94 g_2^4 -\lambda_1\left( 3 g_1^2 +9g_2^2 \right)
 + 12\lambda_1^2 + 4\lambda_3\lambda_4 + 4\lambda_3^2+ 2\lambda_4^2
 + 2\lambda_{1S}^2 \nonumber\\
 &+ 12\lambda_1y_t^2 - 12y_t^4 +\left[0,4\lambda_1 y_\tau^2 - 4y_\tau^4\right] , \\
 \mathcal{D} \lambda_2 = &\ 
   \frac34 g_1^4 + \frac32 g_1^2g_2^2 + \frac94 g_2^4 -\lambda_2\left( 3 g_1^2 +9g_2^2 \right)
 + 12\lambda_2^2 + 4\lambda_3\lambda_4 + 4\lambda_3^2+ 2\lambda_4^2
 + 2\lambda_{2S}^2\nonumber \\
 &+ 12\lambda_1y_b^2 - 12y_b^4 
  +\left[4\lambda_1 y_\tau^2 - 4y_\tau^4,0\right] 
  + 4\lambda_2\text{Tr}\left(y_\nu^\dagger y_\nu\right) -4\text{Tr}\left(y_\nu^\dagger y_\nu y_\nu^\dagger y_\nu\right), \\
 \mathcal{D} \lambda_3 = &\ 
   \frac34 g_1^4 - \frac32 g_1^2g_2^2 + \frac94 g_2^4 -\lambda_3\left( 3 g_1^2 +9g_2^2 \right)
 + \left(6\lambda_3+2\lambda_4\right)\left(\lambda_1+\lambda_2\right) +4\lambda_3^2+2\lambda_4^2
 + 2\lambda_{1S}\lambda_{2S} \nonumber \\
 &+ \lambda_3\left( 6y_t^2+6y_b^2+2y_\tau^2+2\text{Tr}\left(y_\nu^\dagger y_\nu\right) \right)
 - 12y_t^2y_b^2 - \left[0, 4\left(y_\nu^\dagger y_\nu\right)_{33}y_\tau^2 \right] , \\
 \mathcal{D} \lambda_4 = &\ 
   3 g_1^2g_2^2 -\lambda_4\left( 3g_1^2 + 9g_2^2 \right)
 + 2\lambda_4\left(\lambda_1+\lambda_2\right) + 8\lambda_4\lambda_3 + 4\lambda_4^2 \nonumber \\
 &+ \lambda_4\left( 6y_t^2+6y_b^2+2y_\tau^2+2\text{Tr}\left(y_\nu^\dagger y_\nu\right) \right)
 + 12y_t^2y_b^2 + \left[0, 4\left(y_\nu^\dagger y_\nu\right)_{33}y_\tau^2 \right] , \\
\textcolor{blue}{\underline{\mathcal{D} \lambda_S}} = &\ 
   10\lambda_S^2 +2\lambda_S \text{Tr}\left( y_N^\dagger y_N\right)
   + 4\lambda_{1S}^2 + 4\lambda_{2S}^2 -2\text{Tr}\left( y_N^\dagger y_N y_N^\dagger y_N \right), \\
 \mathcal{D} \lambda_{1S} = &\ 
   \lambda_{1S}\left( -\frac32 g_1^2 -\frac92 g_2^2 
 + 4\lambda_{1S} +4\lambda_S +6\lambda_1  \right) + \lambda_{2S}\left(4\lambda_3 + 2\lambda_4\right) \nonumber \\
 &+ \lambda_{1S}\left( 6y_t^2 + \left[ 0 , 2y_\tau^2 \right] + \text{Tr}\left(y_N^\dagger y_N\right) \right) , \\
 \mathcal{D} \lambda_{2S} = &\ 
   \lambda_{2S}\left( -\frac32 g_1^2 -\frac92 g_2^2 
 + 4\lambda_{2S} +4\lambda_S +6\lambda_2  \right) + \lambda_{1S}\left(4\lambda_3 + 2\lambda_4\right)\nonumber\\
 &+ \lambda_{2S}\left( 6y_b^2 + \left[ 2y_\tau^2 , 0 \right] + 2\text{Tr}\left(y_\nu^\dagger y_\nu\right) 
 + \text{Tr}\left(y_N^\dagger y_N\right) \right) 
 - 4\text{Tr}\left(y_\nu^\dagger y_\nu y_N^\dagger y_N\right) , \\
 \mathcal{D}\epsilon = &\ 
 \epsilon \left( -\frac32 g_1^2 -\frac92 g_2^2 
 + 2\lambda_3 + 4\lambda_4 + 2\lambda_S + 4\lambda_{1S} + 4\lambda_{2S}
 + 3y_t^2 + 3y_b^2 + y_\tau^2 + \text{Tr}\left(y_\nu^\dagger y_\nu \right) + \text{Tr}\left(y_N^\dagger y_N \right) \right), \\
 \mathcal{D} y_t = &\ 
   y_t\left( -\frac{17}{12}g_1^2-\frac94 g_2^2-8g_3^2
   +\frac92 y_t^2+\frac12 y_b^2 + \left[0,y_\tau^2\right] \right) , \\
 \mathcal{D} y_b = &\ 
   y_b\left( -\frac{5}{12}g_1^2-\frac94 g_2^2-8g_3^2
   +\frac92 y_b^2+\frac12 y_t^2 + \left[y_\tau^2,0\right] \right) , \\
 \mathcal{D} y_\tau = &\ 
   y_\tau\left( -\frac{15}{4}g_1^2-\frac94g_2^2
   +\frac52 y_\tau^2+ \left[3y_b^2,3y_t^2\right] \right) , \\
\textcolor{blue}{\underline{\mathcal{D}y_\nu}} = &\ 
   y_\nu \left( -\frac34 g_1^2 -\frac94 g_2^2 + 3y_b^2 + \text{Tr}\left(y_\nu^\dagger y_\nu\right) \right)
   + \left[ y_\nu y_\tau^2 -\frac32 \text{Diag}\left(0,0,y_\tau^2\right)y_\nu , 0 \right]
   + \frac32 y_\nu y_\nu^\dagger y_\nu + \frac12 y_\nu y_N^\dagger y_N,  \\
\textcolor{blue}{\underline{\mathcal{D}y_N}} = &\ 
   \frac12 \text{Tr}\left( y_N^\dagger y_N\right) y_N + y_N y_N^\dagger y_N
   + y_N y_\nu^\dagger y_\nu + y_\nu^T y_\nu^\ast y_N .
\end{align}

\pagebreak

\section{Explicit example RG evolution \label{AppRGEPlots}}

Shown below is the RG evolution of dimensionless parameters
in our explicit example (as a function of $\log_{10}\mu_R$).
\\

  \centering
 \includegraphics[width=0.63\textwidth]{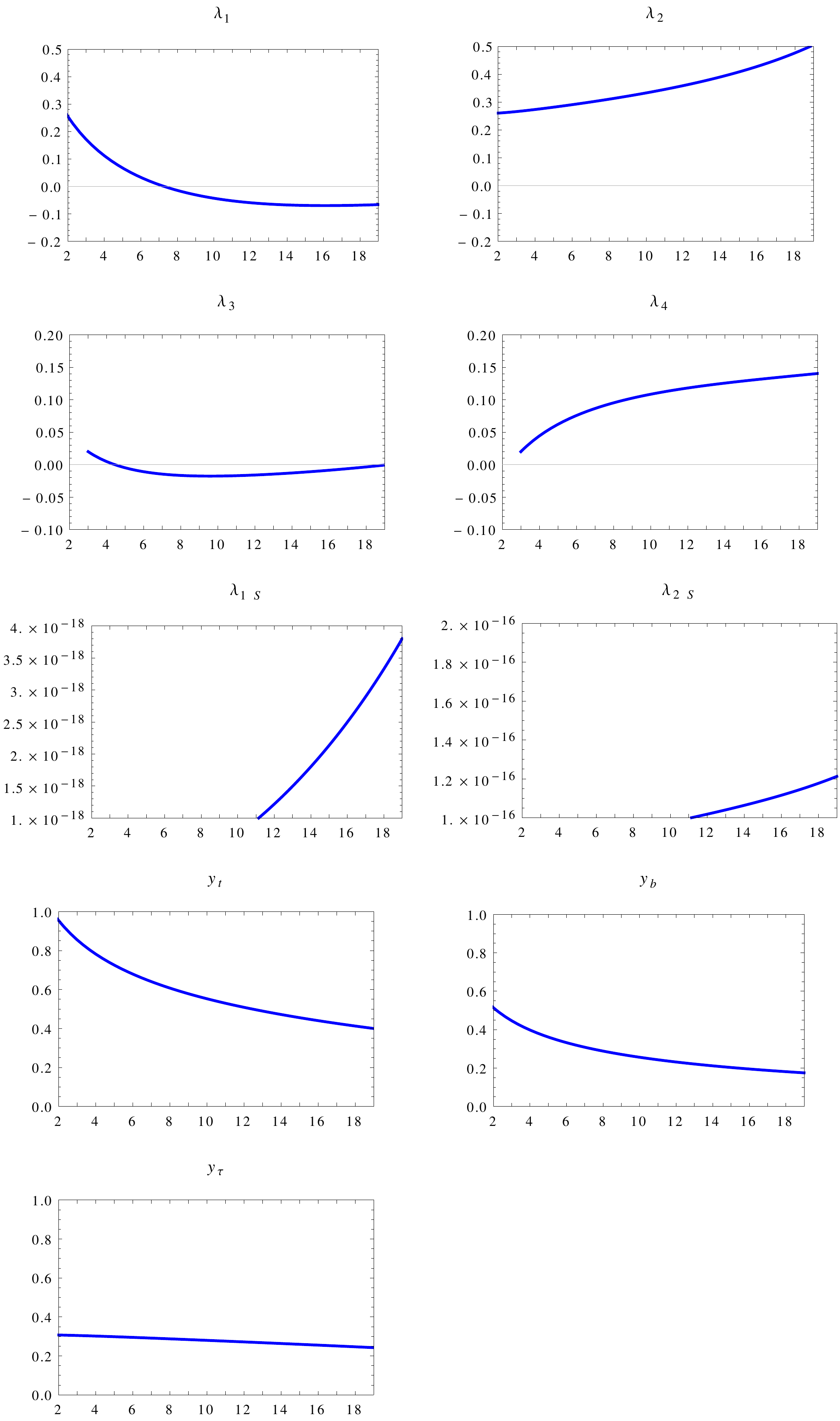}

\end{widetext}

\end{document}